\documentclass[pdflatex,sn-mathphys-num]{sn-jnl}

\usepackage{graphicx}%
\usepackage{multirow}%
\usepackage{amsmath,amssymb,amsfonts}%
\usepackage{amsthm}%
\usepackage{mathrsfs}%
\usepackage[title]{appendix}%
\usepackage{xcolor}%
\usepackage{textcomp}%
\usepackage{manyfoot}%
\usepackage{booktabs}%
\usepackage{algorithm}%
\usepackage{algorithmicx}%
\usepackage{algpseudocode}%
\usepackage{listings}%

\theoremstyle{thmstyleone}%
\theoremstyle{thmstyletwo}%

\theoremstyle{thmstylethree}%

\begin{document}

\title[Article Title]{The role of A in ARPES}

\author*[1]{\fnm{Sergey} \sur{Borisenko}}\email{s.borisenko@ifw-dresden.de}

\affil*[1]{\orgname{IFW-Dresden}, \orgaddress{\street{Helmholtzstrasse 20}, \city{Dresden}, \postcode{01069}, \country{Germany}}}

\abstract{Angle-resolved photoemission spectroscopy (ARPES) was developed in the 1970s–1990s as a momentum-resolved probe of the electronic structure of solids, with its basic principles summarized in early seminal works and classic reviews. The basic ARPES result is currently almost exclusively  associated with the two-dimensional intensity distribution with coordinate axes labeled as \textit{Binding Energy} and \textit{Momentum}. Such datasets are usually directly compared to the band structure calculations along high symmetry directions of the  Brillouin zone. Here we show that this is the result of an oversimplified treatment of the ARPES data leading to incorrect interpretations of the experimental results. We suggest practical ways of how to avoid it in both data collection and analysis.
}

\keywords{ARPES, synchrotron, electronic structure, band structure, energy distribution curve}

\maketitle

\section{Introduction}\label{sec1}

Angle-resolved photoemission spectroscopy has a long history dating back to the 1970s, when the method was first established as a tool for probing the electronic structure of solids with momentum resolution. The fundamental concepts of the photoemission process, angular distributions, and matrix-element effects were developed and discussed in early experimental and theoretical works, later summarized in influential reviews, such as e.g. Ref \cite{Himpsel01011983}. These foundational studies defined the original assumptions, limitations, and native variables of the technique, many of which remain directly relevant for the interpretation of modern ARPES experiments.

From this fundamental perspective, to define a direction in space, you need two angles. The intensity distribution of e.g. radiation or particles emitted into the half-space is conveniently represented by a circular colour map with polar coordinates, where the reference direction is a normal. This is standard practice, which is widespread in numerous scientific approaches. 

Surprisingly, in the case of angle-resolved photoemission spectroscopy (ARPES), despite every tutorial starts with the basic sketch of the experiment showing exactly
these two angles, another unit of information has gained a significant
popularity. Currently more than 90\% of images suggested by the search
engines as a response to the search word \textit{ARPES} are showing the color plots
in Cartesian system with one axis reserved for another variable - binding
energy. This calls for a direct comparison of ARPES data with the band
structure calculations, which a vast majority of ARPES papers do show: in
many cases, down to 5 or even 10 eV binding energies plots are shown next to
or overlapping with band structure along the high symmetry directions with
indication of high-symmetry $k$-points. Another common dataset is a Fermi
surface map, often shown directly compared to the constant 
$k_{z}$-cut through the
calculated Fermi surface (FS).

While we do not doubt or question great achievements of the method as the most informative and effective tool to study electronic structure of solids and most of the conclusions of the previous work will remain valid, we believe the corrections we suggest here will allow to look at the technique from a slightly different perspective beneficial not only for newcomers but also, as we noticed, for quite seasoned ARPES experts.

 As we show below, simple kinematics of the experiment implies that using the single photon energy:
 \begin{itemize}
  \item[--] it is not possible to probe any high-symmetry direction of the Brillouin zone;
  \item[--] widely accepted unit of information is a sophisticated projection in the four-dimensional space, which cannot be regarded as intensity vs binding energy and momentum along any line (direction) in BZ;
  \item[--] it is not possible to probe energy distribution corresponding to a single $k$-point - intensity is recorded only for one energy point, i.e. EDC is in this regard a confusing term.
\end{itemize}
We further suggest the correct way to obtain energy distribution for a given $k$-point and emphasize the advantages of an alternative approach which uses a different unit of information, native to the method.  
 
\section{$E$-approach}\label{sec2}

\begin{figure}[h!]
\centering
\includegraphics[scale=0.2]{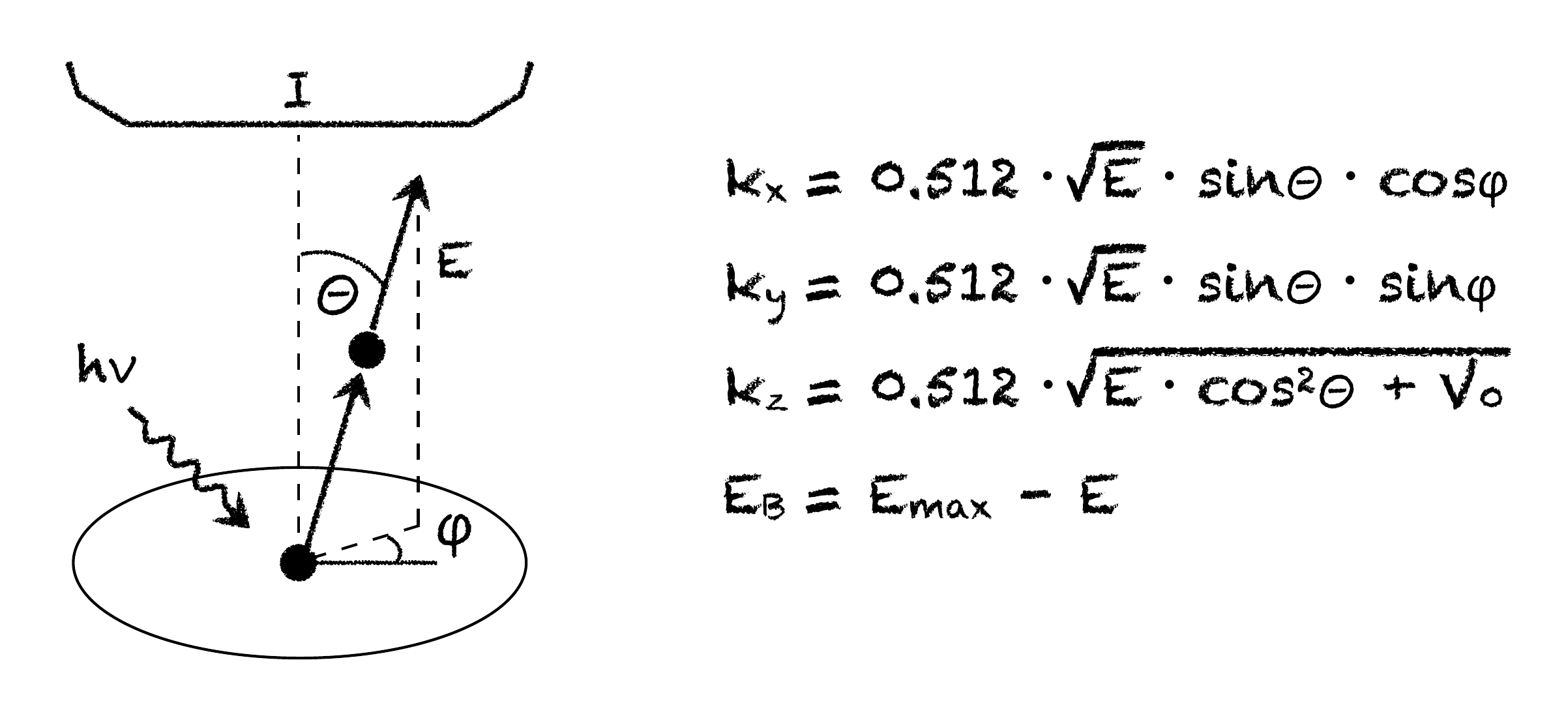}
\caption{Schematic representation of the experiment and basic relationships between the variables. Momentum is given in \AA$^{-1}$, energy - in electronvolts.}\label{one}
\end{figure}

We define the simplest version of the technique as counting the number of photoelectrons as a function of four variables: the two angles mentioned above and the photon and kinetic energies (Fig.~\ref{one}). 
We also assume that parallel component of the final state momentum $k$ is conserved and use energy conservation in its simplest form $E=k^{2}/2m-V_{0}$, where $E$ is a kinetic energy of the photoelectron and $V_{0}$ is a crystal potential \cite{MahanPRL}. It is easy to see from Figure~\ref{one} that photoelectrons with a given $E$ correspond to the sphere in reciprocal space with a radius of $0.512$\AA$^{-1}\sqrt{(V_{0}+E)/eV}$ and therefore this radius would inevitably change if $E$ were to change. In a certain analogy to the sphere of reflection (Ewald's sphere), we will refer to it as the \textit{sphere of emission}.

\begin{figure}[h]
\centering
\includegraphics[scale=0.2]{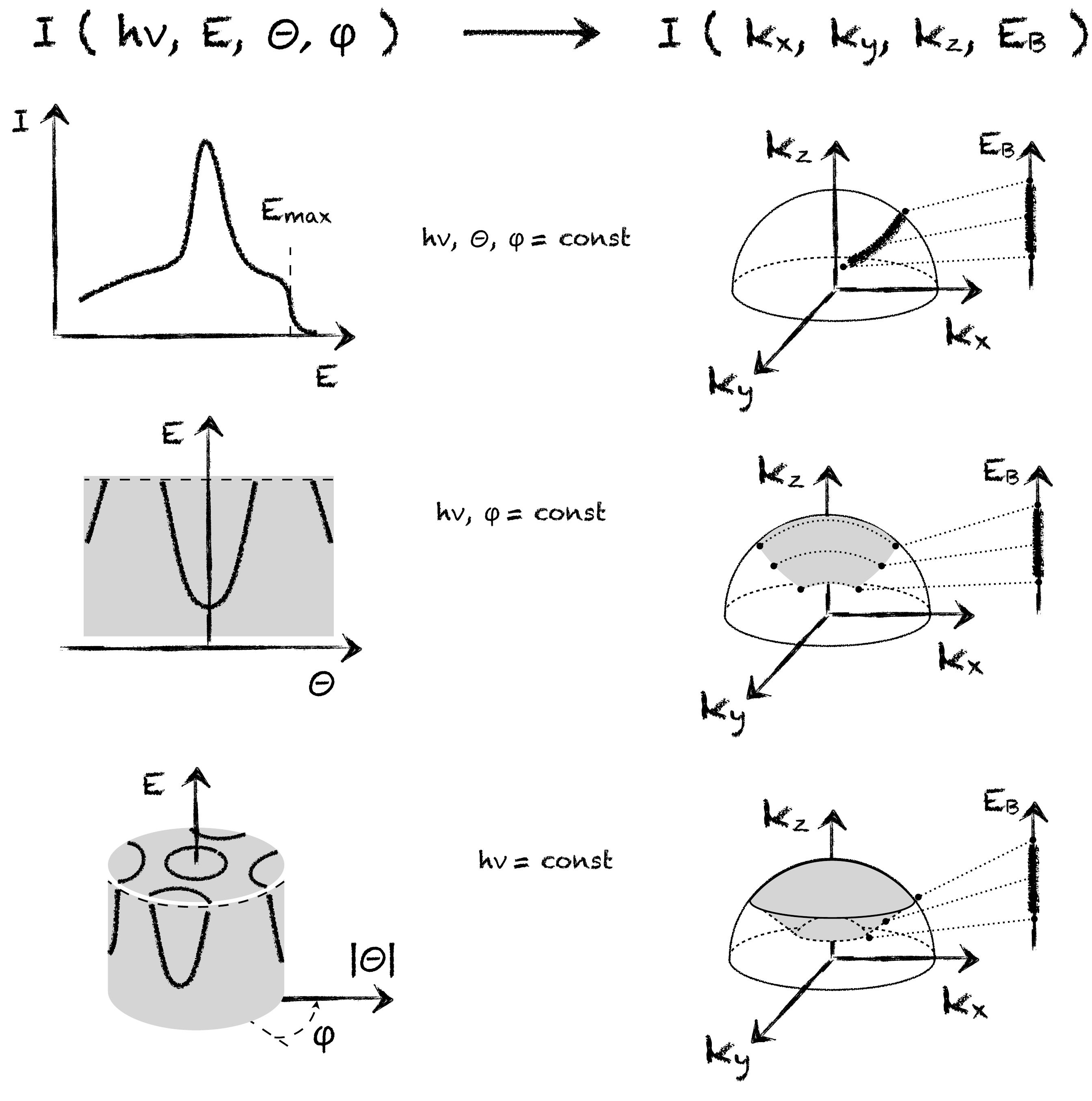}
\caption{E-approach. Left column - datasets currently recorded in ARPES. Right column - corresponding portions of the ($k$, $E_{B}$)-space. The intensity in 2D and 3D diagrams is given by grey color scale. Emission sphere is shown by its contours. Dotted lines indicate that each energy point is uniquely connected to particular momentum.}\label{two}
\end{figure}

In Figure~\ref{two} we show the correspondence between the datasets typically acquired in ARPES and portions of the four-dimensional ($k$, $E_{B}$)-space in this simplest approximation. The first row demonstrates that the spectrum widely known as energy distribution curve (EDC) corresponds to actually both, energy and momentum distributions in ($k$, $E_{B}$)-space. As kinetic energy is reduced, so are all three components $k_{x}$, $k_{y}$ and $k_{z}$ (see also Fig. \ref{one}). This path in $k$-space deviates from radius since $k_{z}$ remains finite when $E$ goes to zero. Even if recorded for normal emission, this curve would not correspond to a single $k$. For this reason, we believe that the term EMDC (\textit{energy-momentum distribution curve)} would be more appropriate for such curves to reflect their non-trivial physical meaning .

The next row shows mentioned above famous ARPES dataset, taken along arbitrary path in the angular space, which in Fig. \ref{two} is chosen by keeping angle $\phi$ constant. We will refer to this dataset as \textit{rainbow-plot} to emphasize both, the shape of the probed portion of the momentum space and the fact that each arc corresponds to always different binding energy (similar to colors of a rainbow). It is now clear, that the validity of the approximation when comparing this dataset with band structure calculations along a high-symmetry direction is defined by proximity of this 2D area in the $k$-space to a 1D straight line. We have not found a term for such part of the circle in the literature and will call it \textit{diffsector} implying that it is close to difference between two circular sectors for circles with different radii.

In the lowest panels of Fig.\ref{two} we show the dataset which is usually collected by additionally varying the other angle while recording multiple rainbow-plots at a fixed photon energy. Although covering a whole spherical diffsector in the $k$-space it still cannot be used to extract true energy distributions for particular $k$-points for the reasons explained above: kinetic energy of photoelectron in vacuum uniquely defines the absolute value of its momentum and via conservation laws - the absolute value of the momentum $k$ of the final state. 
\begin{figure}[h]
\centering
\includegraphics[scale=0.2]{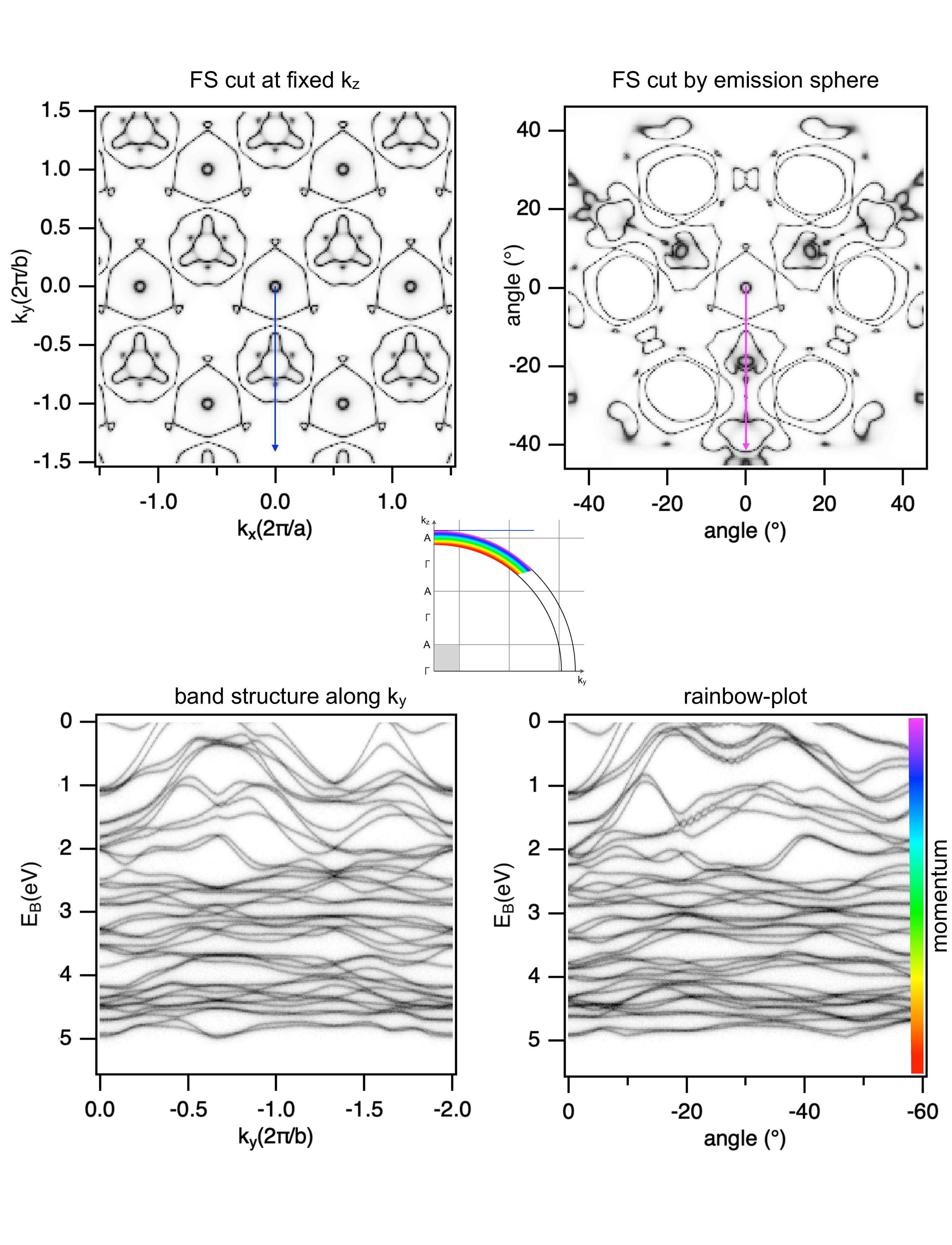}
\caption{Band structure calculations of PtBi$_{2}$. Upper left panel: FS contour at $k_{z}$ approximately corresponding to $\Gamma$A/2. Upper right: FS contour cut by emission sphere of the radius 2.69 \AA$^{-1}$ corresponding to $E_{max}$ = 17.6 eV and $V_{0}$ = 10 eV. Inset: Irreducible part of the BZ projected to $k_{y}k_{z}$ plane has the dimensions 0.96 x 1.02 \AA$^{-1}$ and is marked by the grey shaded area. Rainbow area and colors show that both energy and momentum are not constant. Lower left: band structure along $k_{y}$ as indicated by blue arrow/line above. Lower right: rainbow plot representing band structure along magenta arc}\label{three}
\end{figure}
To quickly estimate the \textit{rainbow} effect quantitatively, we recall that the typical size of the irreducible part of the BZ is about 0.5 \AA$^{-1}$. Usual measurements using HeI-radiation down to 5 eV binding energy simultaneously covers more than 50 \% of this value only in terms of radius of the emission sphere. 

We illustrate another estimate in Fig.\ref{three}, where we consider band structure calculations for a real material (PtBi$_{2}$). Here we assume that $V_{0}$=10 eV and $E_{max}$ is equal to 17.6 eV. These values define the radius of the emission sphere and thus $k_{z}$ for normal emission, at which we cut the FS and show it together with the corresponding energy bands in the left panels of Fig.\ref{three} using a repeating zone scheme. In the right panels we show the results from the same calculations, but for the momentum and energy values probed by ARPES using 22.1 eV photons. Since the material is 3D, the FS changes significantly upon changing $k_{z}$. The difference between the constant-$k_{z}$ cut and Fermi surface map recorded at this photon energy is dramatic. The inset explains which portions of the BZ are probed by recording a rainbow-plot in a wide angular range. Usually ignored variations of momentum are immense, when compared to the size of the irreducible part of the BZ (gray shaded area). All this makes physical meaning of the rainbow-plots very opaque. There are hardly any similarities in the dispersing curves of both plots, apart from the limited region in the upper left corner, where the magenta region runs close to the blue line, as implied by the inset. If the situation is less symmetrical and the magenta-coloured arrow does not lie in the high-symmetry plane, which is almost always the case in real experiments, the chances of generating a corresponding band structure for comparison are even smaller, not to mention its intuitive meaning.

One could imagine that all complications with the unknown and varying $k_{z}$ become automatically irrelevant in the case of ARPES in two-dimensional materials. In this regard we note, that no material is truly two-dimensional since the atom and its electronic orbitals are essentially 3D objects. Moreover, without the 3D-lattice, the "2D-electrons" cannot gain sufficient momentum in perpendicular direction and thus cannot be emitted from the material, i.e. would not be even detected by ARPES.

One way to take into account the mentioned effects is actually shown in Fig.\ref{three} - one can compare ARPES maps with the spherical cuts of the FS and rainbow-plots with the band structure calculated for every arc and every energy. However, especially the latter are far from being intuitive and can hardly quickly contribute to our understanding of the electronic structure of a particular material. Moreover, $V_{0}$ is not apriori known.

Another option is to try to minimize the rainbow-effect and collect the data for materials with small lattice constants, large $V_{0}$ and using higher photon energies, smaller emission angles and smaller energy ranges. In terms of the elements of the inset to Fig.\ref{three} - one should keep the magenta region close to the blue line and keep deviations small comparing to the grey-shaded area.

If we want to compare ARPES results with the usual output of the band structure calculations, i.e. Fermi surfaces and dispersions along high-symmetry directions,
variation of the photon energy becomes inevitable. In order to map the full 3D Fermi surface one would need to record multiple rainbow-plots (and use only small fraction of the recorded data - angular distribution at $E_{max}$) to cover angular space corresponding to irreducible part of the BZ in kx and ky directions at a fixed photon energy and then repeat all these measurements for each of the multiple photon energies to cover irreducible part of the BZ along $k_{z}$. If the lifetime of the sample surface permits, this would be a very (synchrotron beam-) time-consuming experiment. 
Futhermore, shifting the magenta arrow around (Fig. \ref{three}) must involve either rotation of the sample itself or deflecting the electron beam. Both result in anisotropic angular resolution, irregular lattice of probed points even in angular space, additional breaking of cylindrical symmetry, more experimental parameters and require more sophisticated data analysis. Indeed, the traditional abandonment of angular information in favour of entangled momentum-energy mix has led to the fact that, in spite of cylindrically symmetric electron lenses and thus photoelectron trajectories, the sample environment almost never preserves this geometry, natural to the method ( Fig.\ref{one} ). 

There is no other way to record also the energy distribution for each $k$-point, or true EDCs, which are needed to compare with the calculations along high symmetry directions - as relations in Fig.\ref{one} imply, changing photon energy is much more direct and the only approach to probe $E_{b}$ without changing momentum \footnote{Photon energy directly defines $E_{max}$, the maximal kinetic energy of photoelectrons at zero temperature. The electrons, which are always closer to the surface than the nuclei, form a kind of capacitor with the latter, which slows down the photoelectrons and leads to a difference between h$\nu$ and $E_{max}$. This difference is usually of the order of 4-5 eV and is known as work function.}. Only recording the photocurrent as a function of photon energy while keeping both angles \textit{and} kinetic energy constant gives true energy distribution for a given $k$-point. If the true EDC should be as detailed as having e.g. 100 points, one needs to measure using 100 different photon energies. Again, in order to know which exactly $k$-point is probed, one needs to know $V_{0}$. It can be determined if h$\nu$-measurements are  carried out and clear periodicity of the photoemission pattern is detected. Usually the rainbow-dataset centered around normal emission is recorded and not at constant $E$, but keeping $E_{b}$-range constant. Also in this case unnecessary information is collected since if desired energy range is fully probed for desired momentum points the rainbow-plot necessarily provides many incomplete energy distributions for numerous other $k$-points. Finally we note, that in order to identify the normal emission itself, the mapping described above needs to be carried out even prior to h$\nu$-scan.

\section{A-approach}\label{sec3}

\begin{figure}[h]
\centering
\includegraphics[scale=0.2]{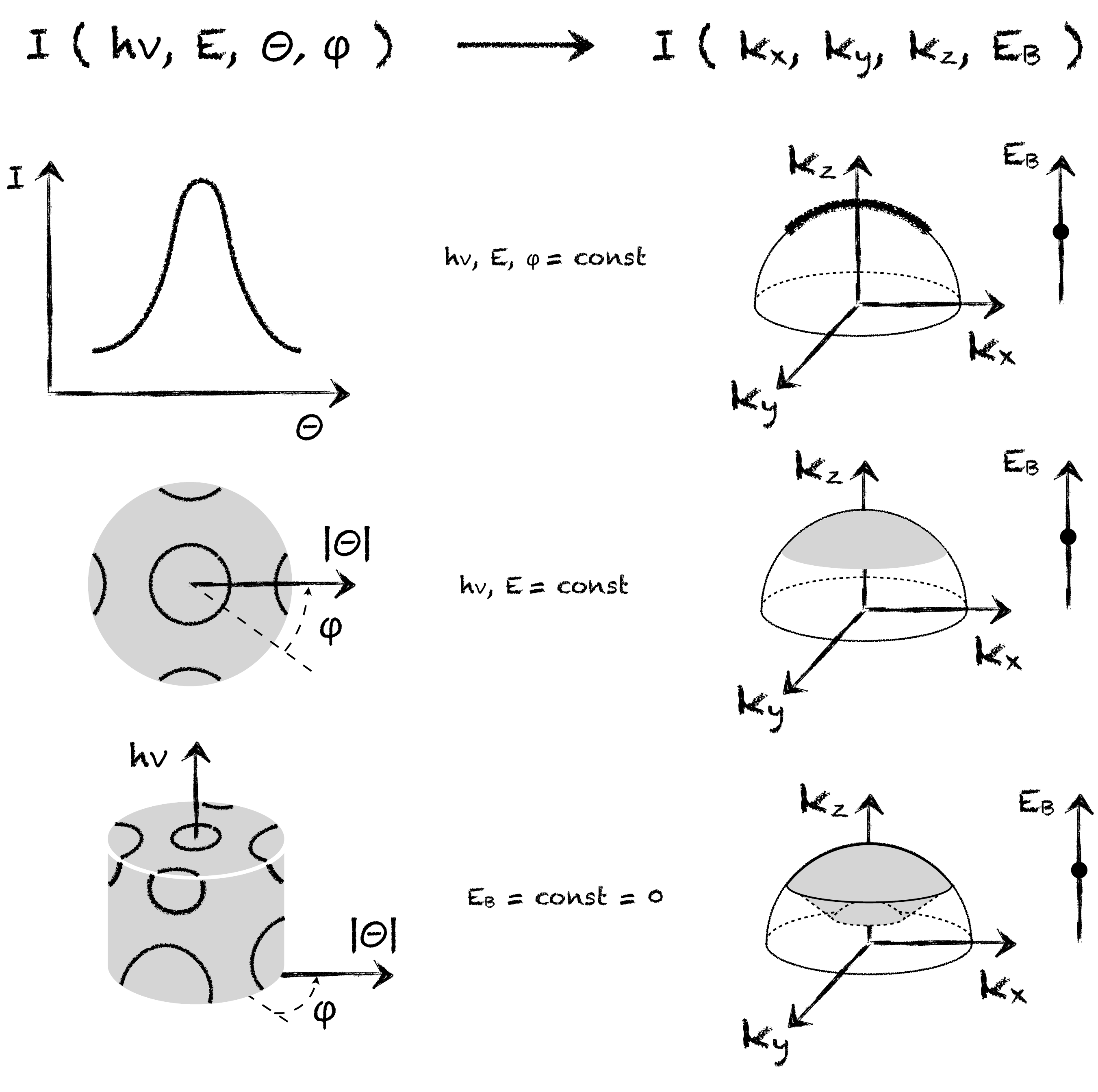}
\caption{Schematics}\label{four}
\end{figure}

We suggest at this point to recall that A in ARPES stands for angle, to shift the focus of the discussion away from the kinetic energy of photoelectrons and to concentrate instead on canonical $\theta$ and $\phi$. Fig.\ref{four} introduces an alternative basic unit of information for ARPES. If we consider a single angular distribution curve obtained by fixing one of the angles and both energies, we immediately obtain dataset with transparent physical meaning - it is an intensity distribution corresponding to an arc path in the BZ (upper panels). 

Perhaps the most natural ARPES dataset is sketched in the second row of Fig.\ref{four}. It is given in polar coordinates and can be obtained by varying also the second angle. Collected in this way $\theta$-$\phi$-\textit{plot} represents the 2D angular distribution of ARPES intensity and if taken at $E_{max}$, it directly corresponds to the section of the 3D Fermi surface by the emission sphere. This distribution can be directly measured in 2D with isotropic angular resolution and regular lattice of sampled points. Already a single dataset is typically sufficient to identify the normal emission direction and simple linear shift in angular space is the only necessary correction of the Fermi surface map before switching to $k$-scale.

Adding another dimension by varying photon energy, one arrives at the
possibility to fully map the 3D Fermi surface in a spherical diffsector. This is illustrated by the sketches in the lower panels of Fig.\ref{four}. 
The example of such measurements can be found in Ref. \cite{Tomography}.

Alternatively, one may wish to keep $E$ (not $E_{B}$, as in the third row of Fig.\ref{four}) constant and record true energy distribution by varying h$\nu$. In this case - for all momentum points covered by $\theta$-$\phi$-plot simultaneously. Note, that in this case every recorded data point is used in the analysis. There are no unnecessary taken data, neither for Fermi surface mapping nor for energy distributions at given $k$-points, contrary to $E$-approach discussed earlier. 

All datasets in this approach are more intuitive also in the sense of a possibility to directly compare with the results of the band structure calculations. Even a small portion of the 3D Fermi surface may provide necessary information about $V_{0}$. Knowing $V_{0}$ one can conveniently navigate in 3D BZ. The proposed way of recording the true energy distributions of the intensity opens up the possibility to compare the band structure practically along any direction in the BZ. However, the curvature of the emission sphere should still be taken into account.

The experiments can now be carried out while maintaining the cylindrical symmetry, which in particular offers the possibility of adjusting the angular acceptance. If it is not sufficient, the sample can be biased and the electrons accelerated while accessing a larger and now symmetrical part of the momentum space. Overcomplicated manipulations of the sample, which inevitably lead to poorer energy resolution and higher base temperatures, become superfluous.

\section{Conclusion}\label{sec3}
The obvious advantages of the intuitive $\theta$-$\phi$- against rainbow-  unit of information is implied already by the Fig.\ref{one}. Kinetic energy enters all four relations, while photon energy, though extending the limits in which $E$ can vary, directly defines only $E_{max}$. It is this decoupling of energy and momentum which is crucial for A-approach.

Surprisingly, if the emission sphere is occasionally taken into account when analyzing the photon energy dependent data, the rainbow-aspect, to our best knowledge, has escaped the papers and tutorials so far and we must admit, also ours. The additive action of both effects, as the inset to Fig.\ref{three} demonstrates, clearly implies that both need to be taken into account when deriving quantitative information, such as bandwidth, large-scale renormalization, Luttinger count, band dispersions etc. Correct comparison with DFT-results may solve a lot of mysteries previously attributed to sophisticated many-body physics. On the other hand, if location in the BZ is well controlled, recording $E$-scans within, let's say 100 meV from the Fermi level, could still be a good approximation and quantities like superconducting gaps or Fermi velocities could still be derived. The routines can be written, similar to the one we used to make the plot in the lower left panel of Fig.\ref{three}, which   
allow a direct comparison. In other words, one can enjoy $E$-scans in the laboratory using a single photon energy source wisely.

We believe that current domination of the $E$-approach is a consequence of the historical development of the method and accessibility of the instrumentation. The early X-ray photoemission spectroscopy dealt almost exclusively with the energy distribution. Clearly, synchrotron beam-time remains very limited and scanning kinetic energy is a relatively simple and well-developed technique, accessible even using laboratory sources delivering single or a few photon energies. That's probably why $E$-scans are extensively used as a substitute for measuring the true energy distribution. Correct EDCs, with the rare exception of reconstructions from h$\nu$-scans in the normal emission, have yet to be measured. 

It is the angular distribution of the outgoing photoelectrons that makes ARPES such a remarkable and powerful method, linking two worlds: the quantum mechanical one, in which the electrons with the same crystal momentum may have different energies, and the classical one, in which the two are clearly linked. 
We hope that the first letter of the famous abbreviation will regain its strength.

\backmatter

\bmhead{Acknowledgements}
We are grateful to Alexander Yaresko for performing band structure calculations and to Volodymyr Zabolotnyy and Mykhailo Varvarin for the help with their presentation. We thank Philip Hofmann, Charlotte Sanders, Timur Kim, Xingjiang Zhou, Konrad Matho, Dan Dessau, Alexander Kordyuk, Simon Moser, Andriy Kuibarov, Oleksandr Suvorov, Susmita Changdar, Alexander Fedorov, Rui Lou and Bernd Büchner for fruitful discussions.

\bibliography{sn-bibliography}

@article{Himpsel01011983,
author = {F.J. Himpsel},
title = {Angle-resolved measurements of the photoemission of electrons in the study of solids},
journal = {Advances in Physics},
volume = {32},
number = {1},
pages = {1--51},
year = {1983},
publisher = {Taylor \& Francis},
doi = {10.1080/00018738300101521},


URL = { 
    
        https://doi.org/10.1080/00018738300101521
    
    

}

}

@article{MahanPRL,
  title = {Angular Dependence of Photoemission in Metals},
  author = {Mahan, G. D.},
  journal = {Phys. Rev. Lett.},
  volume = {24},
  issue = {19},
  pages = {1068--1070},
  numpages = {0},
  year = {1970},
  month = {May},
  publisher = {American Physical Society},
  doi = {10.1103/PhysRevLett.24.1068},
  url = {https://link.aps.org/doi/10.1103/PhysRevLett.24.1068}
}

@article{Tomography,
  author		= "Borisenko, S. and others",
  title			= "Fermi surface tomography",
  journal		= "Nature Communications",
  volume		= "13",
  pages			= "4132",
  year			= "2022",
  doi			= "10.1038/s41467-022-31841-z"
}

@article{ARPES,
  author		= "Borisenko, S.",
  title			= "99 years of {A}{R}{P}{E}{S}",
  journal		= "Nature Reviews Physics",
  volume		= "3",
  pages			= "539",
  year			= "2021",
  doi			= "https://doi.org/10.1038/s42254-021-00350-1"
}

\end{document}